\documentstyle[preprint,eqsecnum,aps,prb]{revtex}
\begin{document}
\title{Distinct origins of magnetic -field -induced resistivity irreversibility in
two manganites with similar ground states : Pr$_{0.5}$Sr$_{0.41}$Ca$_{0.09}$%
MnO$_{3}$ and La$_{0.5}$Ca$_{0.5}$MnO$_{3}$ }
\author{R. Mahendiran, A. Maignan, C. Martin, M. Hervieu, and B. Raveau}
\address{Laboratoire CRISMAT, ISMRA, Universit\'{e} de Caen, 6 Boulevard du\\
Mar\'{e}chal Juin, 14050 Caen-Cedex, France}
\date{\today}
\maketitle

\begin{abstract}
Our investigation of the magnetotransport in two charge ordered manganites
with similar magnetic ground states reveals that the origin of
magnetoresistance can not be concluded from the isofield resistivity, $\rho $%
(T, constant \ H), measurements alone. Both Pr$_{0.5}$Sr$_{0.41}$Ca$_{0.09}$%
MnO$_3$ (PrSrCa) and La$_{0.5}$Ca$_{0.5}$MnO$_3$ (LaCa) show a ferromagnetic
transition (T$_C$ = 260 K for PrSrCa, 230 K for LaCa) followed by an
antiferromagnetic transition (T$_N$ = 170 K for PrSrCa, 140 K for LaCa).
These compounds show qualitatively similar magnetotransport : Below the
irreversibility temperature T$_{IR}$, field cooled (FC) resistivity is lower
than zero field cooled (ZFC) and decreases continuously with T, \ whereas
the ZFC $\rho $(T, H) resembles $\ $the behavior of $\rho $(T, H = 0 T). The
value of $\rho $(ZFC)/$\rho $(FC) is $\approx $ 10$^4$ at 5 K and $\mu _0$H
= 7 T in both compounds. However, isothermal magnetic measurements suggest
distinct origins of magnetoresistance : Field cooling enhances ferromagnetic
phase fraction in LaCa whereas it drives PrSrCa into a metastable state with
high magnetization. The distinct origins of magnetotransport is also
reflected in other magnetic history dependent properties.
\end{abstract}


\section{\protect\smallskip INTRODUCTION}

\smallskip There has been a burst of interest in RE$_{1-x}$AE$_x$MnO$_3$%
-type manganites (RE = La$^{3+}$, Pr$^{3+}$, Nd$^{3+}$ etc., AE = Ca$^{2+}$,
Sr$^{2+}$, Ba$^{2+}$, etc.,) because of their extraordinary sensitivity of
resistivity ($\rho $) to internal molecular and external magnetic fields.
The $\rho $(T) \ evolution in general shows a metallic like behavior (d$\rho 
$/dT 
%TCIMACRO{\TEXTsymbol{>} }
%BeginExpansion
\mbox{$>$}%
%EndExpansion
0) in the ferromagnetic (FM) state and an insulating like behavior (d$\rho $%
/dT 
%TCIMACRO{\TEXTsymbol{<} }
%BeginExpansion
\mbox{$<$}%
%EndExpansion
0) in the antiferromagnetic (AF) state.\cite{Rao} The essential physics of
electrical transport in manganites is understood in terms of the Double
Exchange (DE) interaction which allows spin conserving transfer of e$_g$
hole along Mn$^{3+}$:t$_{2g}^3$e$_g^1$-O$^{2-}$ -Mn$^{4+}$:t$_{2g}^3$e$_g^0$
network if the core t$_{2g}^3$ spins are aligned ferromagnetically and
forbids this hole exchange if the t$_{2g}^3$spins are antiparallel.\cite
{Zener} In addition, A-site ionic radius (%
%TCIMACRO{\TEXTsymbol{<}}
%BeginExpansion
\mbox{$<$}%
%EndExpansion
r$_A$%
%TCIMACRO{\TEXTsymbol{>}}
%BeginExpansion
\mbox{$>$}%
%EndExpansion
) and size mismatch ($\sigma ^2$) of the A-site cations, orbital and charge
ordering, and electron-phonon coupling can enhance the magnitude of the
resistivity.\cite{Rao,Attfield,Millis} A large decrease of $\rho $(T) under
an external magnetic field occurs close to the magnetic phase boundary: the
ferromagnetic-paramagnetic phase boundary shifts up and the charge ordered,
antiferromagnetic-ferromagnetic or paramagnetic phase boundary shifts down
in temperature with increasing field.\cite{Rao} There is no clear consensus
about the origin of magnetoresistance (MR) although the basic mechanism can
be traced to the field induced ferromagnetic alignment of t$_{2g}^3$ spins
and the resulting delocalization of the e$_g$-carriers. Structural
transition, electron-spin-phonon coupling, melting of charge and orbital
order, and phase separation which all can be magnetically tunable appear to
play an important role in determining the magnitude of the magnetoresistance.%
\cite{Asamitsu}

\bigskip

The magnitude of \ MR is path dependent in some of the manganites: \ MR is
lower when the sample is cooled in zero field (ZFC) prior to establishing a
magnetic field at low temperature than when cooled\ in a field (FC) from
high temperature. Only some investigators have payed attention to this
aspect.\cite{Xiao,Tokura,Hervieu,Huang} However, its origin is not
understood yet. Although the temperature T$_{IR}$ below which the ZFC and FC 
$\rho $(T) bifurcate generally shifts down in temperature with increasing H,
an exceptional case of upward shift of T$_{IR}$ with H was also found in Pr$%
_{0.5}$Ca$_{0.5}$Mn$_{0.98}$Cr$_{0.02}$O$_{3\text{.}}$\cite{Mahi1} The most
elaborate study under different values of magnetic field was carried out by
Xiao et al. \cite{Xiao} on La$_{0.5}$Ca$_{0.5}$MnO$_3$. The ZFC and FC
measurements reported so far were done in an isofield temperature scan mode,
i.e., temperature dependence of \ $\rho $ under a field, but this method
masks various contributions to the irreversibility. Hence, we investigated
electrical and magnetic properties of La$_{0.5}$Ca$_{0.5}$MnO$_3$ (LaCa) and
Pr$_{0.5}$Sr$_{0.41}$Ca$_{0.09}$MnO$_3$ (PrSrCa) in detail. Both of these
compounds have similar magnetic properties: ferromagnetic transition at high
temperature (T$_C$ = 240 K for LaCa, 265 K for PrSrCa) and charge ordering
and antiferromagnetic transition at low temperature (T$_N$ = 170 K, T$_N$ =
140 K $\leq $ T$_{CO}$ for LaCa, T$_N$ = T$_{CO}$ = 170 K for PrSrCa). The $%
\rho $(T) of PrSrCa compound is metallic like (d$\rho $/dT 
%TCIMACRO{\TEXTsymbol{>} }
%BeginExpansion
\mbox{$>$}%
%EndExpansion
0) in between T$_C$ and T$_N$ whereas it is insulating in LaCa. We will show
that the origin of the resistivity irreversibility in these two magnetically
identical compounds are different although they look the same from an
isofield temperature scan.

\section{\protect\bigskip EXPERIMENT}

The polycrystalline Pr$_{0.5}$Sr$_{0.41}$Ca$_{0.09}$MnO$_3$ (PrSrCa) and La$%
_{0.5}$Ca$_{0.5}$MnO$_3$ (LaCa)samples were prepared by the standard ceramic
route and were characterized by neutron diffraction, electron diffraction
and transport \ measurements.\cite{Damay,Mahi2} We remeasured the four probe
resistivity of polycrystalline PrSrCa and LaCa using Quantum Design made
Physical Property Measuring System in different field modes: In the zero
field cooled (ZFC) and field cooled (FC) modes, $\rho $(T) was measured
while warming from 5 K in a field H1 after cooling in H = 0 and H = H1
respectively. Resistivity isotherms at 5 K in ZFC and FC were also measured.
In the FC(7T)-FW(H) mode, the sample was cooled from T = 300 K to 5 K in a
high field of 7 T, then the field was reduced to a lower H value at 5 K and
data were taken while warming in H. In the FC(H)-FW(0T) mode, the sample was
cooled from T = 300 K to 5 K under a high field, the field was reduced to
zero at 5 K and data were taken while warming in zero field. Ac
susceptibility in H$_{ac}$ = 1 Oe and f = 100 Hz was also measured in 300
K-5 K temperature range.

\section{\protect\bigskip RESULTS}

Fig. 1(a) and Fig. 1(b) show the temperature dependence of $\rho $(T, H = 0
T) of PrSrCa, and LaCa upon cooling and warming. The insets shows the real
part ($\chi $') of the ac susceptibility. The ferromagnetic and
antiferromagnetic transitions are marked. The $\rho $(T, 0) of PrSrCa
exhibits a metallic like behavior in between T$_C$ and T$_N$ and it jumps
abruptly at T$_N$ (= 170 K) while cooling. The value of $\rho $ increases by
7 orders of magnitude in between T$_N$ and 5 K. The $\rho $(T, 0) decreases
rapidly as T$_N$ is approached from below and T$_N$ shifts up to 180 K upon
warming. The charge and antiferromagnetic ordering in PrSrCa occur at the
same temperature.\cite{Damay} A pronounced hysteresis in $\rho $(T, 0) and $%
\chi $'(T) occur around the Neel temperature. The $\rho $(T, 0) behavior of
LaCa is similar to that reported by other groups.\cite{Roy} In contrast to
PrSrCa, $\rho $(T, 0) of LaCa does not exhibit a metallic behavior in
between T$_C$ and T$_N$.which is otherwise expected according to double
exchange mechanism\cite{Zener}. The $\rho $(T, 0) increases abruptly around
140 K while cooling and increases by more than 6 orders of magnitude in
between 140 K and 5 K. Below 30 K, $\rho $(T, 0) becomes asymptotic to the
temperature axis as seen previously.\cite{Roy} While warming $\rho $(T, 0)
exhibits a hysteresis over a wide temperature range (90 K- 220 K) and also
is the susceptibility (see the inset in Fig. 1(b)). Although the charge and
antiferromagnetic orderings in LaCa are generally assumed to occur
simultaneously\cite{Goodeneough}, our results to be presented latter
indicate that charge ordered clusters are present already at 220 K (i.e., in
the ferromagnetic regions) even {\it while cooling} from 300 K.

\bigskip

Fig. 1(c) and Fig. 1(d) show $\rho $(T, H) in the ZFC and FC modes for
PrSrCa and LaCa respectively. We can clearly see an irreversibility between
the FC and ZFC curve for all field value. The irreversibility starts at a
temperature (T$_{IR}$) close but below T$_N$.\ The T$_{IR}$ is determined
from the temperature [$\rho $(ZFC)-$\rho $(FC)]/$\rho $(ZFC) $\leq $ 3 \%. \
T$_{IR}$ = 160 K (155 K), 150 (150 K), 133 K (149 K) for $\mu _0$H = 2 T, 4
T, 7 T for PrSrCa (LaCa)]. Below T$_{IR}$, the ZFC- $\rho $(T,H ) continues
to rise and reaches a higher value than the FC- $\rho $(T) which \ either
ascends or descends with decreasing T depending upon the field strength. The
temperature dependence of the ZFC and FC curves of PrSrCa and LaCa are
qualitatively similar. The ZFC-$\rho $(T,H)) in both compounds mimics the
temperature dependence of $\rho $(T, 0) curve albeit. At 5 K and at 7 T the
resistivity ratio $\rho $(ZFC)/$\rho $(FC) = $\approx $ 7.5 x 10$^4$ for
PrSrCa is comparable to $\rho $(ZFC)/$\rho $(FC) $\approx $ 3.8 x 10$^4$ for
LaCa. Thus, from the isofield measurement alone it appears that a single
mechanism is responsible for the irreversibility behavior.

\bigskip

In order to understand the origin of the irreversibility in details, we
studied the field dependence of the resistivity, and the magnetization.
Figures 2(a) and (b) show isothermal scans for PrSrCa and LaCa respectively
at T = 5 K. When H is increased from 0 T to 7 T in the \ zero field cooled
mode, $\rho $(T =5 K, H) of PrSrCa decreases rather smoothly by almost 3
orders of magnitude and when H is reduced to zero $\rho $(5 K, H) does not
attain the original starting value. Upon further cycling \ from 0 T$%
\longrightarrow $ -7 T $\longrightarrow $0 T, $\rho $ exhibits hysteresis
but the loop is closed. The field cycling process in the FC mode is H $%
\rightarrow $0 T $\rightarrow $-H $\rightarrow $0 T $\rightarrow $H. The
field cooled resistivity at any vlaue of the field is clearly lower than the
corresponding zero field cooled value.\ The hysteresis loop is open in the
positive field cycle with lower resistivity in \ H $\rightarrow $0 T branch
than 0 T $\rightarrow $ H branch. In the ZFC mode, $\rho $(5K, H) of LaCa
also decreases rather smoothly up to the maximum field of $\mu _0$H = 7 T.
But the field dependence (see the curvature) is different from LaCa. It also
exhibits hysteresis like the PrSrCa compound but the loop is not closed when
H is increased from -5 T to 0 T. . When the sample is field cooled, the
hysteresis is negligible compared to PrSrCa.\bigskip

Figures 2(c) and 2(d) compare M(H) of PrSrCa and LaCa respectively. In the
ZFC mode , M(H) of PrSrCa shows a weak ferromagnetic like behavior for $\mu
_0$H 
%TCIMACRO{\TEXTsymbol{<} }
%BeginExpansion
\mbox{$<$}%
%EndExpansion
\ 0.5 T and then increases with a constant slope up to the maximum field of
5 T. The maximum magnetic moment at 5 T in the ZFC mode is 0.185 $\mu _B$
which is only 5.3 \% of the maximum moment of 3.5 $\mu _B$ expected if the
whole sample were ferromagnetic \ When the sample is field cooled under 5 T,
M at 5 T increases to 0.483 $\mu _B$ which is nearly 2.6 times larger than
the corresponding field value in the ZFC mode. The M(H) loop is symmetric
about the origin when the field is cycled between -5 T and +5 T after the
virgin 5 T $\rightarrow $0 T curve. But, the value at\ $\mu _0$H = 5 T
decreases to 0.3 $\mu _B$ when the field is increased from -5 T to +0.5 T.
We have carried out M(H) under FC mode for $\mu _0$H = 4 T, and 3 T. Since
the behaviors are qualitatively the same as 5 T data we do not show them
here. The M(H) of LaCa (Fig. 2d) is in many sense contrasting to PrSrCa. The
ZFC curve exhibits a ferromagnetic like behavior but with a large moment at
0.5 T. The value M at 0.5 T is 0.18 $\mu _B$ for LaCa whereas it is only
0.04 $\mu _B$ for PrSrCa. At 5 T, M of LaCa is 0.38 $\mu _B$ which is twice
as large as PrSrCa. A small hysteresis is seen only in the positive field
cycle. When LaCa is field cooled under 5 T, M at 5 T increases to 0.85 $\mu
_B$ however, unlike in PrSrCa, M does not reduce to low value at 5 T after \
5T $\rightarrow $0T$\rightarrow $--5T$\rightarrow $5T cycle. More
importantly we see a large enhancement of \ the low field magnetization in
LaCa compared to PrSrCa. The spontaneous magnetic moment M$_0$ = 0.68 $\mu
_B $ (determined by extrapolating the linear high field part to H = 0 )
decreases with the strength of the cooling field. The presence of \ a
ferromagnetic phase below the charge ordering temperature is in agreement
with the results inferred from other techniques\cite{Roy,Mori} and supports
the phase separation scenario proposed theroretically.\cite{Dagotto} The
volume fraction of the ferromagnetic phase in LaCa can be calculated from f$%
_m$ = M$_0$/M$_S$ where M$_S$ = 3.5 $\mu _B$ and M$_0$ is the spontaneous
magnetic moment. The f$_m$ \ in LaCa increases from 4.95 \% in \ the ZFC
mode to 19.69 \% when \ field cooled in 5 T. These are only rough estimates
without considering canting of moments in the antiferromagnetic phase. When
extrapolated from the recently published data on Pr$_{0.5}$Sr$_{0.5-x}$Ca$_x$%
MnO$_3$ ( x = 0.1, 0.2)\cite{Neibi}, the low field behavior M(H) in our
PrSrCa compound can be ascribed to\ few ten nanometer size ferromagnetic
clusters (f$_m$ 
%TCIMACRO{\TEXTsymbol{<} }
%BeginExpansion
\mbox{$<$}%
%EndExpansion
0.8 \%) which can behave like a superparamagnetic entity. These observations
underline distinct origins of the path dependent magnetoresistance in these
two magnetically identical compounds.

\bigskip

The above presented zero field cooled data in both compounds do not indicate
any evidence of a field induced antiferromagnetic-ferromagnetic
(metamagnetic) transition up to 5 T which is otherwise expected. To know
whether we are dealing in the field range far below the critical field for
metamagnetic transition, we carried out M(H) measurement at different
temperatures and for higher field values ($\mu _{0}$H 
%TCIMACRO{\TEXTsymbol{>} }
%BeginExpansion
\mbox{$>$}%
%EndExpansion
5 T) in particularly for PrSrCa. Magnetization up to 12 T for PrSrCa sample
was measured using a vibrating sample magnetometer at University of
Zaragoza, Spain. Fig. 3 (a) shows M(H) data for PrSrCa. The measurements
were done in zero field cooled mode. The M(H) curve at 210 K is a typical of
a long range ferromagnet. But for lower temperatures, the M(H) , for example
at T = 150 K, initially increases linearly but then jumps abruptly to a
higher value around the threshold field H$_{C}$ = 4 T and at H 
%TCIMACRO{\TEXTsymbol{>}}
%BeginExpansion
\mbox{$>$}%
%EndExpansion
%TCIMACRO{\TEXTsymbol{>} }
%BeginExpansion
\mbox{$>$}%
%EndExpansion
H$_{C}$, M(H) appears to saturate. The transition is first order as can be
guessed from the hysteretic behavior. The abrupt jump characterize the
metamagnetic transition from the antiferromagnetic state to either a spin
flop or a spin flip state. For H 
%TCIMACRO{\TEXTsymbol{>}}
%BeginExpansion
\mbox{$>$}%
%EndExpansion
%TCIMACRO{\TEXTsymbol{>} }
%BeginExpansion
\mbox{$>$}%
%EndExpansion
H$_{C}$, the sample is ferromagnetic and for H 
%TCIMACRO{\TEXTsymbol{<} }
%BeginExpansion
\mbox{$<$}%
%EndExpansion
H$_{C}$, the increase of M(H) is due to canting of antiferromagnetic moment
away form the direction of \ the spin axis towards the field direction. The
critical field for the metamagnetic transition increases from 4 T at 150 K
to 8 T at 75 K and to more than 11 T at 25 K.

\bigskip

\ We also carried out\ M(H) for LaCa \ but the field was restricted to the
maximum available field of 5 T in the SQUID magnetometer at Laboratoire
CRISMAT, Caen, but \ the available field range (0 T -5 T) is sufficient for
the purpose of this paper. The results are shown in Fig. 3(b). At T = 150 K $%
\approx $ T$_{N}$, M(H) behaves like a ferromagnet until $\mu _{0}$H = 2 T,
but then \ increases rapidly above 3.5 T due to the metamagnetic transition.
The metamagnetic transition is not completed at 5 T. At 5 T, the magnetic
state can be characterized as a mixture of \ magnetic field induced
ferromagnetic domains \ (in addition to ferromagnetic phase present in zero
field) and antiferromagnetic domains with canted moments. From the
spontaneous magnetization M$_{0}$ = 0.6 $\mu _{B}$ we calculate the volume
fraction of the ferromagnetic phase f$_{m}$ = 17.15 \%. At 125 K, f$_{m}$
decreases to $\approx $ 5.7 \% and $\approx $ 4.95 \% at 5 K. The M(H) curve
of LaCa at 5 K is more like ferromagnet because the metamagnetic transition
takes place at much higher fields.

\bigskip

Surprisingly, M(H) data at \ T = 175 K which is above T$_N$ = 140 K but
below T$_C$ = 230 K also show a metamagnetic transition above $\mu _0$H = 1
T with hysteresis. However, the metamagnetic transition is not as sharp at
150 K or as in PrSrCa compound. The M(H) data at 200 K also closely resemble
the 175 K but with a very small hysteresis. We suggest that charge and
orbital ordered domains exist above the Neel temperature in contrast to the
assumption that charge ordering and antiferromagnetic orderings occur
simultaneously.\cite{Goodenough} The metamagnetic transition in this
temperature range is caused by the field induced destruction of the charge
and orbital order. The delocalization of e$_g$- carriers enhances
ferromagnetic ordering of t$_{2g}^3$ spins and so the magnetization
increases. The broad metamagnetic transition suggests that domains of
variable sizes are present. Some indirect evidence for charge ordering in
the ferromagnetic region can be quoted from the very recent literature: A
large positive volume magnetostriction above T$_N$ in La$_{0.5}$Ca$_{0.5}$MnO%
$_3$ was reported earlier by us\cite{Mahi2} and it was suggested that charge
ordered phase of lower unit cell volume coexist with \ a ferromagnetic phase
of higher unit cell volume in between T$_C$ and T$_N$. The ferromagnetic
phase having higher unit cell volume than the charge ordered phase was
recently confirmed by Huang et al.\cite{Huang} ( see Fig. 7 in ref. 10) who
also found formation of a new structural phase around 220 K in La$_{0.5}$Ca$%
_{0.5}$MnO$_3$. The new structural phase is most likely charge and orbital
ordered although Huang et al.\cite{Huang} did not explicitly mention it. The
charge-orbital domains are most likely paramagnetic\cite{Rivadulla} in this
temperature range since no antiferromagnetic signal was detected in between
220 K and 150 K.\cite{Huang} As the temperature is lowered the charge
ordered domains grow in size coalesce around T = 140 K below\ which $\rho $%
(T = 0) starts to increase rapidly.\ A long range antiferromagnetic ordering
also takes place around 140 K. A question remains to be unanswered is why
the resistivity does not show metallic behavior in between T$_C$ and T$_N$.
A possibility is that charge ordered domains above T$_N$ but below T$_C$ are
sandwiched between ferromagnetic domains and hence impedes charge transfer
between ferromagnetic domains. However, we can not arrive any definitive
conclusion from the present study alone and future work should address this
issue.

\bigskip

\section{Discussion}

Now we turn our attention again towards the origin of \ the field induced
irreversibility in resistivity. The magnetization study suggests that the
ferromagnetic phase fraction at 5 K in LaCa (f$_m$ $\approx $ 4.5 \%) is
larger than in PrSrCa (fm $\approx $0.8 \%). It is likely that the
ferromagnetic phase is concentrated in some regions of the sample in LaCa
and is randomly distributed in PrSrCa. The difference in the curvature of
the virgin ZFC -$\rho $(H) curves below 2 T in Fig. 2(a) and 2(b) suggests
such a possibility. When the sample is zero field cooled, the magnetization
of \ the ferromagnetic domains is randomly oriented in LaCa and as H is
increased from 0 T to 2 T, after domain wall motion, ferromagnetic domains
reorient in the field direction. This causes a rapid decrease of the
resistivity in LaCa in this field range.\cite{Roy,Mori} The spin canting of
the antiferromagnetic sublattices at higher fields (H 
%TCIMACRO{\TEXTsymbol{>} }
%BeginExpansion
\mbox{$>$}%
%EndExpansion
2 T) can also contributes to the magnetization. The main source of the
magnetoresistance in this field regime in the zero field is tunnelling of \
the spin polarized carriers between the ferromagnetic domains and between
ferromagnetic domains and canted antiferromagnetic phase. In PrSrCa, the
dominant contribution to MR in the zero field mode is the canting of
antiferromagnetic sublattice. The ZFC behavior of $\rho $(H) at 5 K in both
PrCa \ and LaCa suggests that metamagnetic transition does not take place up
to 7 T. Field cooling has two effects. First, it gives a \ preferential
orientation of M of ferromagnetic domains which already exist in LaCa.
Second, new ferromagnetic domains are formed at high temperature due to the
destruction of the charge ordered domains. The second contribution is more
effective in LaCa than in PrSrCa because charge ordered domains of various
sizes and hence different critical fields (H$_C$) are already present above
the Neel temperature in LaCa. This new ferromagnetic phase continues to
exist down to the lowest temperature in LaCa. Hence, the ferromagnetic phase
fraction increases with the strength of the cooling field in LaCa and so, $%
\rho $ is lowered in the field cooled mode. If the ferromagnetic domains
form a percolating network (H $\geq $ 2 T ), not necessarily to be a 3
dimensional network but can be filamentary\cite{Dagotto}, then the
temperature dependence of $\rho $ is mainly determined by the thermal
dependence of the magnetization of the ferromagnetic phase and so $\rho $(T)
increases with increasing temperature. There are two possible scenarios in
the case of PrSrCa. It appears that field cooling does not enhance the
ferromagnetic phase fraction in PrSrCa if one interprets the magnetization
data (Fig. 2(a)). So, the first possibility is that the high field state is
metastable with\ a majority antiferromagnetic domains with nonzero net
magnetisations coexisting with nanometric ferromagnetic clusters. In the
double exchange picture\cite{Zener}, the magnetization due to canting
between two neighboring Mn$^{3+}$ and Mn$^{4+}$ sites is M $=$M$_s\cos
(\theta /2)$ \ where $\theta $ is the canting angle and M$_s$ is the
saturation magnetization. From the observed value of M = 0.483 $\mu _B$ at 5
T and at 5 K in the virgin FC curve (see Fig. 2(c)), we calculate $\theta /2$
\ = 82.06$^{\circ }$ between the antiferromagnetically coupled sublattice.
which is reduced from $\theta /2$ \ = 90$^{\circ }$ in the uncanted
scenario. After field cycling M reduces to 0.2903 $\mu _B$, and the canting
angle increases to $\theta /2$ \ = 85.25$^{\circ }$and in the zero field
cooled mode the canting angle is still higher ($\theta /2$ \ = 86.97$^{\circ
}$). Thus, the low resistivity in the field cooled mode in PrSrCa \ is due
to charge transport through the canted antiferromagnetic domains rather than
by the increase of ferromagnetic phase fraction. Another possibility is that
field cooling (H $\geq $ H$_C$) partially destroys the antiferromagnetic and
charge ordering and creates\ bigger ferromagnetic clusters and
antiferromagnetic, charge ordered domains. The low resistivity in the field
cooled mode can be understood as the result of the percolation of the
ferromagnetic clusters. When the field is reduced from 5 T, the
magnetization of ferromagnetic clusters becomes random and hence the
spontaneous magnetization is not apparent in the M(H) curve. From the
magnetization measurement alone it is not possible to distinguish between
the above two possibilities.

\bigskip\ 

These subtle differences between these two compounds are also reflected in
other measurements. Figures. 4(a) and 4(b) compare the $\rho $(T) behaviors
of PrSrCa and LaCa, respectively, in the FC (H)-FW(0T) mode. The samples
were first cooled rapidly from 300 K to 5 K in a field H, H was reduced to
zero rapidly (250 Oe/sec) and data were taken while warming in zero field.
It can be seen that after reducing H from 7 T to 0 T, the value of $\rho $
at 5 K in PrSrCa is more than 3 orders of magnitude higher than in LaCa.
This reflects the fact that the field cooled state has a longer relaxation
time (more stable) in LaCa than in PrSrCa which is in agreement with the low
field magnetic behavior. When LaCa is heated from 5 K, the FC (7 T)-FW(0T)
curve exhibits a maximum at T$_{max}$ = 75 K, below this temperature, $\rho $%
(T) keeps 'memory' of the low resistive state at 7 T and lost its 'memory'
above 75 K. The T$_{max}$ shifts down (T$_{max}$ = 75 K, 70, 64 K for H = 7
T, 4 T and 2 T respectively) with decreasing H. The FC(7T)-FW(0T) curve of
PrSrCa exhibits a broad maximum at a higher temperature T$_{max}$ = 111 K
and we do not find a clear shift of T$_{max}$ with H. In contrast to LaCa,
the FC(2 T)-FW(0 T) curve does not exhibit a maximum. Figures 4(c) and 4(d)
compare the $\rho $(T) behavior respectively for PrSrCa and LaCa in the FC(7
T)-FW(H) mode. In the FC(7 T)-FW(H) mode, the sample was cooled each time
from 300 K to 5K at a constant field of H = 7 T, the field was reduced to a
H value at 5 K and then $\rho $(T) in the field (H) was measured for
different values of H. \ So for each set of curves presented, the initial
state of the sample at 5 K is the same, \ but a new configuration of domains
and/or canted state is reached by reducing the field from 7 T to the
prescribed H. The resistivity values at 5 K for diffferent H values remain
nearly the same for LaCa but not for PrSrCa which shows a factor of 100
increase between the FC(7 T)-FW(7 T) and the FC(7 T)-FW(1 T) data. The FC(7
T)-FW(H) curve for each H value, except for the FC(7T)-FW(7T) curve in LaCa,
shows a maximum at T$_{max}$ which shifts down continuously with decreasing
H for LaCa but there is a tendency to shift up below 5 T in PrSrCa. In
addition, $\rho $(T) suddenly jumps from a low to high value at the
temperature T$_J$ 
%TCIMACRO{\TEXTsymbol{<}}
%BeginExpansion
\mbox{$<$}%
%EndExpansion
%TCIMACRO{\TEXTsymbol{<} }
%BeginExpansion
\mbox{$<$}%
%EndExpansion
T$_{max}$. The jump is sharper in LaCa, than in PrSrCa, (for example see the
FW(1T) curve). We suggest that the low resistivity below T$_J$ (= 12 K in
LaCa and 60 K in PrSrCa for FW(1T)) is caused by the 'memory' of the low
resistive state at FC (7 T). The downward shift of T$_{max}$ in LaCa can be
understood as follows : The ferromagnetic phase fraction decreases with
decrease in H and hence percolation occurs at a lower temperature.. However,
the dominant effect of reducing H from 7 T in PrCa is the decrease in the
canting angle. The peak in the FC(7 T)-FC(H) occurs when the loss of the
Zeeman energy due to increase in canting angles is compensated by an
increase in thermal energy. Presently no standard formula is available to
fit these data to extract a quantitative information. Nevertheless, these
data indicate that the origins of magnetoresistance in these two compounds
are different.

\bigskip

\section{Summary}

In summary, we have shown that magnetotransport in two magnetically
identical compounds La$_{0.5}$Ca$_{0.5}$MnO$_3$ (LaCa) and Pr$_{0.5}$Sr$%
_{0.41}$Ca$_{0.09}$MnO$_3$ (PrSrCa) are strongly path dependent. Field
cooling is more effective in lowering the resistivity than zero field
cooling. The isofield temperature scan data are qualitatively similar in
both compounds. However, from the isofield temperature scan alone it is
impossible to understand the origin of the magnetic history dependent
irreversibility in resistivity. From the isothermal magnetic measurements,
we have shown that the dominant effect of field cooling is to increase the
ferromagnetic phase fraction in LaCa whereas it creates metastable canted
antiferromagnetic domains and/or ferromagnetic clusters in PrSrCa. If LaCa
is cooled under the maximum field of 7 T, and H is reduced to a different
value at 5 K, the resistivity value is nearly the same for different values
of H at 5 K whereas it increases by two orders of magnitude in PrSrCa. The
resistivity of LaCa when warmed in a field after field cooling in $\mu _0$H
= 7 T, shows a monotonic downward shift of the position of the resistivity
peak but PrSrCa shows a different behavior. These two compounds also exhibit
different resistivity behaviors if samples are heated in zero field after
field cooling in different H. A quantitative understanding of these results
are still lacking. In view of these results, it will be interesting to study
the irreversibility behavior of the resistivity in other half doped
manganites like Pr$_{0.5}$Sr$_{0.5}$MnO$_3$ and Nd$_{0.5}$Sr$_{0.5}$MnO$_3$.
The future work should also address whether the irreversibility in the
resistivity can be observed in a non phase separated manganite.

\bigskip

\section{Acknowledgments}

\bigskip

\bigskip R. M thanks MENRT (France) for financial assistance and thanks
Professor M. R. Ibarra for allowing us to use the vibrating sample
magnetometer for the high field magnetic measurement on one of the samples.

\newpage

\begin{center}
{\bf Figure captions}
\end{center}

\begin{description}
\item[FIG. 1]  Temperature dependence of resistivity at H = 0 T for (a) Pr$%
_{0.5}$Sr$_{0.41}$Ca$_{0.09}$MnO$_3$ (PrSrCa) and (b) La$_{0.5}$Ca$_{0.5}$MnO%
$_3$ (LaCa). Insets show the real part of the ac susceptibility. $\rho $(T)
in zero field cooled (ZFC) and field cooled (FC) modes while warming for (c)
PrSrCa (d) LaCa.

\item[FIG. 2]  $\rho $(H) isotherms at 5 K in the ZFC and FC modes for
different cooling field in (a) PrSrCa (b) LaCa. M(H) isotherms at 5 K in the
ZFC and FC modes in (a) PrSrCa, (b) LaCa.

\item[FIG. 3]  M(H) isotherms for (a) PrSrCa (b) LaCa. The curves were
recorded after cooling \ the sample \ from 300 K to a prescribed temperature
in zero field.

\item[FIG. 4]  $\rho $(T) measured in zero field while warming after cooling
in a field (H) for (a) PrSrCa and (b) LaCa. $\rho $(T) measured in a field H
while warming after cooling in 7 T field for (c) PrSrCa and (d) LaCa.

\newpage 
\end{description}

\end{document}